\documentstyle[twocolumn,aps,pre,graphicx,amssymb,amsbsy,amsfonts]{revtex}

\begin{document}

\draft

\title{SYNCHRONIZATION OF KAUFFMAN NETWORKS}

\author{Luis G. Morelli and Dami\'an H. Zanette}

\address{Consejo Nacional de
Investigaciones Cient\'{\i}ficas y T\'ecnicas\\ Centro At\'omico
Bariloche and Instituto Balseiro, 8400 Bariloche, R\'{\i}o Negro, 
Argentina}

\date{\today}
\maketitle

\begin{abstract}
We  analyze the  synchronization  transition  for a  pair of   coupled
identical Kauffman networks  in the chaotic  phase. The annealed model
for Kauffman networks shows   that synchronization appears   through a
transcritical bifurcation, and provides an approximate description for
the  whole  dynamics  of  the coupled networks.   We show  that  these
analytical predictions are  in  good agreement with numerical  results
for sufficiently   large networks, and   study finite-size  effects in
detail.   Preliminary  analytical and numerical results  for partially
disordered networks are also presented.
\end{abstract}


\section{Introduction} \label{s1}

Synchronization of coupled elements  is a form of collective evolution
present in a variety of complex real  systems and mathematical models.
This class   of emergent  behavior  has  been observed   in biological
populations \cite{winfree}, chemical   reactions     \cite{khrustova},
neural  networks    \cite{golomb},   and  human    social    phenomena
\cite{applause},   among  other  instances.   Models  that account for
synchronization consider, for example,  globally coupled logistic maps
\cite{logmaps},   chaotic  oscillators   \cite{rossler},   Hamiltonian
systems \cite{hamilton}, and formal neural networks \cite{neural}.

In the    usual formulation, two  identical   dynamical  systems whose
individual dynamics is governed by the equation $\dot{\bf w} = {\bf F}
({\bf w})$ are coupled to each other in the form
\begin{equation} \label{introd}
\dot{{\bf w}}_{1,2} = 
{\bf F} ({\bf w}_{1,2}) + 
{\epsilon} \left( {\bf w}_{2,1} - 
{\bf w}_{1,2} \right),
\end{equation}
where $\epsilon$   is  the coupling  parameter.  Full  synchronization
takes  place when  both systems converge   asymptotically to  a common
trajectory, ${\bf   w}_1(t) =   {\bf w}_2(t)$.  When    the individual
dynamics is chaotic---a particularly relevant  case in connection with
the description of real systems---full  synchronization occurs above a
critical  value $\epsilon_c$ of the  coupling intensity. This critical
point is determined by the competition between chaos and coupling, and
can be calculated in terms of the Lyapunov  exponent of the individual
dynamics \cite{logmaps}.

While globally coupled chaotic elements with  a few internal variables
have  been  extensively studied, synchronization of spatially extended
systems remains  quite unexplored. Recently, synchronization  has been
reported for  a     system  consisting   of  two   coupled     complex
Ginzburg-Landau equations  \cite{amengual}. Globally  coupled   neural
networks   \cite{neural},   stochastically coupled  cellular  automata
\cite{eca,bagnoli,grass},   and nonidentical complex   Ginzburg-Landau
systems   \cite{batistutacitadinoonorario},   are  other  examples  of
spatially extended  systems  that present   a critical  transition  to
synchronization.

In this paper  we study  the synchronization  dynamics of  two coupled
identical  Kauffman  networks, which are  discrete  extended dynamical
systems with  quenched disorder.   With respect  to previous  work  on
coupled extended systems, the interest of Kauffman networks resides in
the fact that the transition  to synchronization admits an  analytical
description which results to be in  excellent agreement with numerical
simulations. In Sect.  \ref{kauffman} we briefly review the definition
of Kauffman  networks and  the annealed  model  for the calculation of
their overlaps.  Next,   in  Sect.   \ref{coupling}, we  introduce   a
stochastic coupling  mechanism  for Kauffman networks   and propose an
analytical approach in  the framework  of  the annealed model,   which
identifies  the transition  to   synchronization in  our system   as a
transcritical bifurcation.   Section \ref{numerical}, where  we report
our numerical results, is  the core of  the present paper.   There, we
study the effects of spurious synchronization in finite-size networks,
consider the  application of  noise to the  system to  eliminate  such
effects,  and compare  the   results with the  analytical description.
Remanent finite-size  effects  are  numerically  quantified  and their
analytical treatment, which requires a formulation beyond the annealed
model, is outlined. In Sect.  \ref{dca} we discuss the synchronization
transition in  some  subclasses  of Kauffman  networks, which   may be
thought  of as interpolations  between  generic Kauffman networks  and
ordered  cellular  automata.  Finally,  in   Sect. \ref{summary},  our
results are summarized and discussed.

\section{Kauffman networks} \label{kauffman}

Kauffman  networks,   also  known as   random  Boolean networks,  were
introduced as  a   model  for the    problem  of cell  differentiation
\cite{k1,kauf}. Since then, they have  been the object of many studies
concerning their properties \cite{kauf,d+p,d+w,fly,bp}.

A  Kauffman   network  (KN) is a  disordered   deterministic dynamical
system.  It   consists of  an  $N$-site  network, where  each  site is
connected to $K$ randomly chosen sites.  The parameter $K$ is known as
the \textit{connectivity} of the network.  We refer to  the set of $K$
sites connected  to  a given site as   its \textit{neighborhood}.  The
state of each  site is given by   a Boolean variable  $\sigma _{i} \in
\left\{  0,1 \right\} $,  and evolves according   to the inputs coming
from the neighbor sites.  The  evolution rule is chosen  independently
and randomly for  each site.   To each  possible configuration  of the
neighborhood---there are  $2^{K}$  such configurations---an  output is
assigned, namely,  $1$ with probability $p$,   or $0$ with probability
$1-p$.  The parameter $p\in [ 0,1 ] $ is known as the \textit{bias} of
the rule.  Then,   for each variable   $\sigma _{i}$  we have  Boolean
functions    $f_{i}$  such        that   $\sigma _{i}(t+1)=f_{i}\left[
\boldsymbol{\nu}_{i}(t) \right]  $,  where  $\boldsymbol{\nu}_{i}    =
\left(\sigma_{i_{1}},\dots, \sigma _{i_{K}}   \right) $ is the set  of
inputs of site $i$.  The state  of all sites is updated simultaneously
according to the corresponding  functions.  We can write  an evolution
equation for the state  vector  of the network $\boldsymbol{\sigma}  =
\left( \sigma_{1},\sigma _{2},\dots,\sigma_{N} \right) $, as
\begin{equation} \label{Phi}
{\boldsymbol{\sigma}} (t+1) = {\bf f} 
\left[ \boldsymbol{\sigma} (t) \right] ,
\end{equation}
with
$
{\bf f} \left[ \boldsymbol{\sigma} (t) \right] = 
\left( f_{1} \left[ \boldsymbol{\nu}_{1}(t) \right] ,
f_{2} \left[ \boldsymbol{\nu}_{2}(t) \right] ,\dots,
f_{N} \left[ \boldsymbol{\nu}_{N}(t) \right] \right)
$.

The $K$ connections and the evolution rule  of each site are chosen at
the beginning and kept fixed during the evolution.  Thus, the disorder
is quenched and the dynamics is deterministic.  For a finite number of
sites  $N$, the number of  states in phase  space  is also finite---it
equals $2^N$. Then,   for  any  initial  condition, the  system   will
eventually fall into a cycle.

In  the $(p,K)$ parameter  space, Kauffman networks  present phases of
frozen and chaotic   evolution, separated by   a critical  line.   The
transition between these phases    has been extensively   studied, and
characterized  by  means  of  several  order parameters, such  as  the
Hamming distance \cite{d+p,d+w}  and the stable core size  \cite{fly}.
In most of  this work we  will deal with  the case $p=1/2$ and  $K=3$,
which lies within the chaotic phase.

The annealed   model (AM) was   introduced to  study the evolution  of
overlaps between states in KNs \cite{d+p,d+w}.  In this model, the $K$
connections $\left\{ i_{1},\dots,i_{K}\right\} $ of  each site as well
as the  Boolean  functions $f_{i}$ are randomly   changed at each time
step.  This means   that  an entirely   different realization of   the
network is used   at each step.   Note  that, while  ordinary  KNs are
deterministic, the annealed model  works as a probabilistic automaton.
The asymptotic  periodic behavior  of  KNs is  absent in  the annealed
model.  The advantage  of this model is  that it allows for analytical
calculations,  and it has been shown  that its predictions are in good
agreement with the behavior of KNs in the limit $N \rightarrow \infty$
\cite{d+w}.

Suppose that we  have two identical KNs with  the same connections and
rules.  We feed them  with different initial  conditions, and let them
evolve in time.   We define  the  \textit{overlap} $a(t)$  between the
networks as  the fraction  of homologous sites  that  are in the  same
state at  time $t$.  In  the AM, it is possible  to calculate the time
evolution of the overlap.  At time $t+1$ the connections and the rules
$f_{i}$   are reassigned, but  the  same changes  are  applied to both
networks, keeping them identical. The  probability  for a site  having
all its inputs coming from sites in the same state in both networks is
$a(t)^{K}$. At the next time step,  consequently, such site will be in
the same state  in both networks, no  matter the evolution rule chosen
for it.   Thus,  there is  a fraction  $a(t)^{K}$ of  homologous sites
whose state will coincide at $t+1$.   The remaining $1-a(t)^{K}$ sites
still have a  probability of overlapping.   Even  if the state of  the
neighborhoods of  a given site are  different in the  two networks, it
may happen  that the evolution  rule assigns the  same output to them.
The   probabilities for  $f_{i}\left( \boldsymbol {\nu}_{i}^{1}\right)
=f_{i}\left( \boldsymbol {\nu}_{i}^{2}\right) =0$ and for $f_{i}\left(
\boldsymbol     {\nu}_{i}^{1}\right)     =f_{i}\left(      \boldsymbol
{\nu}_{i}^{2}\right) =1$ are,  respectively,  $(1-p)^2$ and  $p^{2}$.
The overlap at time $t+1$ is then
\begin{equation} \label{evola}
a(t+1) = a(t)^{K}+\left[ 1-a(t)^{K}\right] \left[ p^2+( 1-p )
^2 \right].
\end{equation}

An alternative way to characterize the difference between two networks
is the difference automaton, defined by
\begin{equation} \label{difer0}
d_{i}(t)=\sigma _{i}^{1}(t)\oplus \sigma _{i}^{2}(t) ,
\end{equation}
where $\oplus$ denotes Boolean addition. The density of this automaton 
is given by 
\begin{equation} \label{difer} 
D(t)=\frac{1}{N}\sum_{i=1}^{N}d_{i}(t) ,
\end{equation}
and  coincides  with the Hamming  distance  between the networks. Note
that $D(t)=1-a(t)$ so that, from Eq. (\ref{evola}), we get
\begin{equation}
D(t+1)=2p(1-p)\left( 1-[ 1-D(t)]^{K}\right).  
\label{dda}
\end{equation}
The Hamming distance has  proven to be  a suitable order  parameter in
the study  of  the synchronization  transition in  coupled  elementary
cellular  automata \cite{eca,bagnoli,grass},   where  the analysis  of
overlaps between   states is a  basic  tool to define   the effects of
coupling. In  the next section,  we  adapt the  annealed model to  the
description of coupled KNs.

\section{Coupled Kauffman networks} \label{coupling}

\subsection{Stochastic coupling}

In order  to establish a coupling mechanism  between two KNs, we first
observe that,   due  to   the discrete   nature  of  KNs,   the  usual
deterministic coupling used for maps  \cite{logmaps} cannot be applied
here. Consequently, we introduce a form of stochastic coupling between
networks as previously done  for  cellular automata \cite{eca},  where
the   continuous parameter $q$  that    controls the strength of   the
coupling is a probability, as explained in the following.

The evolution  of the coupled  system is implemented by the successive
application of two operators. First, the evolution function $\bf f$ is
applied  to  both  networks  as if   they  were not   coupled [see Eq.
(\ref{Phi})], yielding ${\bf f} \left( \boldsymbol{\sigma}^{1} \right)
$ and ${\bf  f} \left( \boldsymbol{\sigma}^{2}  \right)  $.  Next, the
stochastic coupling operator ${\cal S}$ is applied:
\begin{equation}
\left\{
\boldsymbol{\sigma}^{1}(t+1),\boldsymbol{\sigma}^{2}(t+1)\right\} =
{\cal  S}  \left( {\bf f} \left[  \boldsymbol{\sigma}^{1}(t)
\right] , {\bf f} \left[ \boldsymbol{\sigma}^{2}(t)  \right]
\right).
\end{equation}
The operator ${\cal  S}$ compares the  states of the networks site  by
site.  If $\sigma_{i}^{1}(t)=\sigma_{i}^{2}(t)$  the state of the site
is not  modified.  If, on  the   other hand, $\sigma_{i}^{1}(t)   \neq
\sigma_{i}^{2}(t)$, with probability $q$ the  states of the homologous
sites in both networks are set to the same value. This value is chosen
among $\sigma   _{i}^{1}(t)$  and $\sigma_{i}^{2}(t)$, with   the same
probability  $1/2$ for each instance. With  probability $1-q$, even if
$\sigma _{i}^{1} (t)\neq \sigma _{i}^{2}(t)$,   the coupling does  not
act,  leaving the state  of that site  unchanged in  both networks. We
call $q$ the  \textit{coupling probability}.  The  whole evolution can
be formally expressed as
\begin{equation}
\left\{ \sigma_i^1 , \sigma_i^2 \right\} \to
\left\{
\begin{array}{ll}
\left\{ f_i \left( \boldsymbol{\nu}_i^1 \right), 
f_i \left( \boldsymbol{\nu}_i^2 \right) \right\} 
& \mbox{with probability } 1-q,\\
\left\{ f_i \left( \boldsymbol{\nu}_i^1 \right), 
f_i \left( \boldsymbol{\nu}_i^1 \right) \right\} 
& \mbox{with probability } q/2,\\
\left\{ f_i \left( \boldsymbol{\nu}_i^2 \right), 
f_i \left( \boldsymbol{\nu}_i^2 \right) \right\} 
& \mbox{with probability } q/2.\\
\end{array}
\right.
\end{equation}

\begin{figure}
\centering
\resizebox{\columnwidth}{!}{\includegraphics[angle=0]{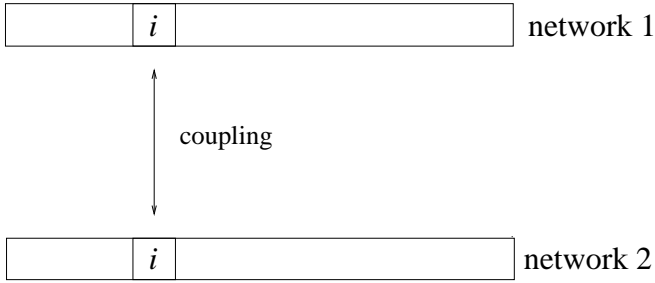}}
\vspace{0.5cm}
\caption{Schematic representation of  the coupling mechanism. Coupling
links homologous elements of  two extended systems---in this case, two
networks. Since  coupling is symmetric,  each  network may act on  the
other.}
\label{acop}
\end{figure}

We stress that we are dealing with two identical extended systems, and
that the coupling mechanism  connects homologous elements of these two
systems, namely, the $i$th site of network 1  is connected by coupling
with the $i$th site of network 2, as schematically illustrated in Fig.
\ref{acop}.  The coupling  mechanism defined above is symmetric, since
each network  may influence the other.  It  could also be  possible to
consider a biased, non symmetric coupling, in which one network drives
the other \cite{bagnoli}.

For  $q=0$ the two networks   are uncoupled, and evolve  independently
from each other.   For  $q=1$, in  contrast,  the networks synchronize
completely at the first time step. From  then on, they follow a common
trajectory in phase space without further intervention of the coupling
mechanism.  Our aim  in the following is  to study the behavior of 
coupled KNs for    intermediate values of the  coupling   probability,
$q\in( 0,1)$.

In  the frozen  phase,  were  no   damage  spreading takes   place, an
arbitrary  small  coupling    intensity  $q>0$   leads   eventually to
synchronization.    The situation    is  different   in  the   chaotic
phase.   There, we find  two  competing driving  forces acting  on the
coupled  system,  namely, the  chaotic    dynamics which induces   the
separation between  two   trajectories to  grow  \cite{kauf},  and the
coupling,  by  which    the  Hamming distance between   the   networks
decreases. In this paper, we focus attention on the chaotic phase.

\subsection{Annealed model for coupled networks}

The annealed model can be used to predict the behavior  of the pair of
coupled KNs.  We  recall that  the time   evolution  equation for  the
Hamming distance in  the case of two {\it  free} networks is  given by
Eq.  (\ref{dda}). Now suppose that  the  Hamming distance  of two {\it
coupled} KNs at time $t$ is $D(t)$.  The first substep in the dynamics
of this system consists  of the free evolution of  both networks.  The
Hamming distance after this  substep, $D( t  +\delta t)$, is therefore
given by Eq.  (\ref{dda}).   At the second  substep, coupling acts  on
the system, and a  fraction $q$ of the  homologous sites that were  in
different states are assigned the same value.   This leaves a fraction
$(1-q) D(  t  +\delta t)$ of sites   with different states  in the two
networks.   Thus, the evolution   of the Hamming  distance for coupled
networks is given by the map
\begin{equation} 
D(t+1)=  F\left[ D(t) \right]=
\varphi (p,q) \left( 1- \left[ 1- D(t) \right]^K \right),
\label{ddac}
\end{equation}
with $\varphi (p,q)=2p(1-p)(1-q)$.

It can be shown that the map (\ref{ddac})  has a stable fixed point 
$D^*>0$ for $q<q_c$, with
\begin{equation}  \label{qc}
q_c = 1 - \left[ 2 p (1-p) K \right]^{-1}.
\end{equation}
At  $q=q_c$    the   system undergoes  a    transcritical  bifurcation
\cite{transcritical},  and $D^* = 0$  becomes a stable fixed point for
$q>q_c$.  Thus,   within the  AM  approximation,  $q_c$ stands for the
critical coupling at  which synchronization sets on.  In this paper we
deal mostly  with the  case $K=3$,  for  which the  stable equilibrium
$D^*$    can  be given analytically  as    a function  of the coupling
probability $q$. In  this case, in  fact, the  map  is defined by  the
cubic function $F(x)=  \varphi (p,q)  x \left(  3 -  3x + x^2  \right)
$. The stable Hamming distance is
\begin{equation} \label{dqam}
D^*(q)= \left\{
\begin{array}{ll}
{3 \over 2} - 
\frac{1}{2} \left[ -3+ \frac{4}{\varphi(p,q)} \right] ^{1/2} 
& \mbox{for }  q < q_c, \\ \\
0 & \mbox{for } q \geq q_c.
\end{array}
\right.
\end{equation}
Note that near the critical point, $q\lesssim q_c$, this
Hamming distance is approximately given by
\begin{equation} 
D^*(q) = 6 p(1-p) |q-q_c|.
\end{equation}
Therefore, the corresponding critical exponent is equal to unity.

For     $q\neq q_c$,   the   Hamming   distance   approaches  $D^*(q)$
exponentially    in  time.   For    $q=q_c$,  on   the   other   hand,
Eq. (\ref{ddac}) can  be approximately written,  for $D(t)  \to 0$, as
$D(t+1)  = D(t)-(K-1) D(t)^2/2$.  This implies  a power-law decay  for
long times, $D(t) \sim t^{-1}$. In the  next section, we compare these
analytical results with those of extensive numerical simulations.
 
\section{Numerical results} \label{numerical}

We have performed numerical simulations  of pairs of KNs coupled under
the  scheme presented above.  The   results  reported in this  section
correspond to the case of $p=1/2$ and $K=3$. We have recorded the time
evolution  of  the  Hamming  distance,  performing averages   over $r$
realizations of $N$-site networks.   The   number of realizations   is
chosen in such  a  way that,  for different values   of $N$, $rN  \geq
10^6$.  In a typical realization, we start with two identical networks
with different random initial  conditions.  For each  realization, new
connections and local functions are  chosen.  The networks are allowed
to evolve freely, without coupling,  for  a transient time $\tau$  of,
typically, $10^3$ steps.  This is done for the networks to reach their
asymptotic  dynamics before  coupling is allowed   to act.  After this
transient, we turn the coupling on, reset the  time to zero, and start
measuring $D(t)$.

\begin{figure}
\centering
\resizebox{\columnwidth}{!}{\includegraphics[angle=-90]{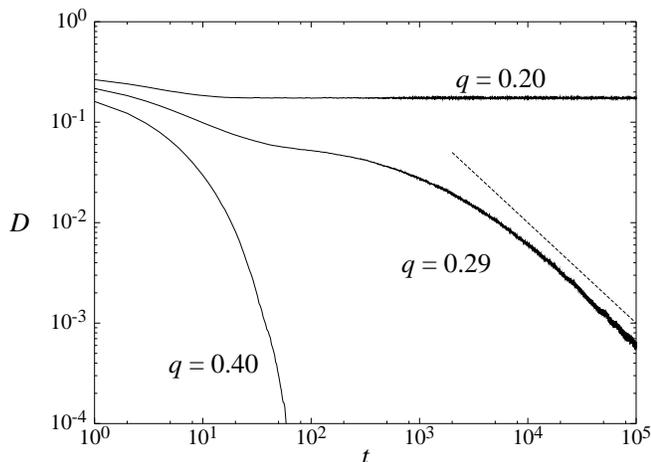}}
\caption{Hamming distance   as a   function of time   for  $10^3$-site
networks and three  coupling  probabilities $q$, averaged over  $10^4$
realizations. The dashed line, of  slope $-1$, is  to be compared with
the power-law decay observed near the critical coupling.}
\label{dt1}
\end{figure}

In Fig.  \ref{dt1} we show the time  evolution of the Hamming distance
$D(t)$ for different values of the coupling parameter $q$.  The values
of   $q$ have been   chosen to display   the  three typical behaviors,
namely,  synchronization for $q > q_c$,  critical decay for $q \approx
q_c$, and convergence to a finite distance for $q < q_c$.

\begin{figure}
\centering
\resizebox{\columnwidth}{!}{\includegraphics[angle=-90]{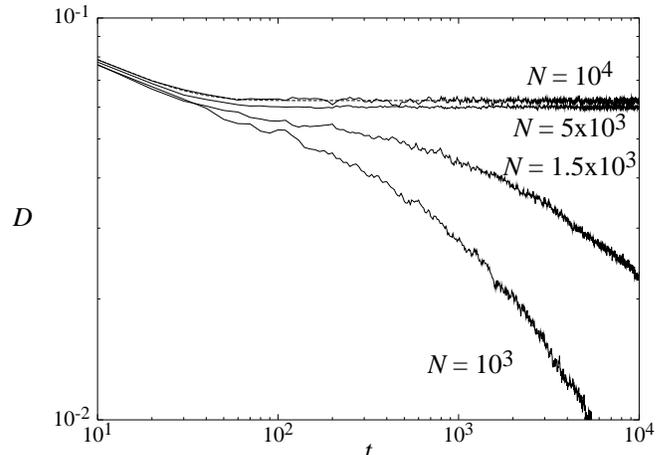}}
\caption{Hamming  distance as a function  of  time for several network
sizes $N$,  and  fixed coupling probability $q=0.29$.    Averages were
done   over $10^3$    realizations for  $N<10^4$,    and over   $10^2$
realizations  for $N=10^4$. The dashed  curve  stands for the annealed
model prediction.}
\label{dt2}
\end{figure}

For the present values of $p$  and $K$, the  annealed model predicts a
critical coupling probability $q_c =1/3$ [cf.  Eq. (\ref{qc})].  It is
however clear  from Fig. \ref{dt1} that  the power-law decay in $D(t)$
is observed for a lower coupling, $q  =0.29$.  Simulations of the same
networks   with  $q=1/3$,  on  the    other   hand, always  lead    to
synchronization.  In fact, the annealed model is expected to provide a
good approximation   to   our system  in   the limit  $N  \to  \infty$
\cite{d+w}.  Figure \ref{dt2} shows the time  evolution of the Hamming
distance   for a fixed  coupling  probability,  $q=0.29$, and  several
values of $N$.    The AM prediction    from Eq. (\ref{ddac})  is  also
shown.  The  strong  dependence with  the   size  of the   network  is
apparent.   In particular, we find   that  for this coupling  strength
$10^3$-site networks synchronize  whereas $10^4$-site networks do not.
The AM result gives a good description for the case of $N=10^4$.

The dependence of $D(t)$  on the size of  the system can be  partially
ascribed to the effect  of {\it spurious synchronization}.  Because of
the discrete nature of KNs, two finite-size networks can be brought to
the same state by a fluctuation caused by  the stochastic coupling. In
such  case, the two    networks  will remain  synchronized from   then
on. This event is more frequent for small networks, where the relative
size of fluctuations increases.  Near the critical point, furthermore,
where the Hamming   distance vanishes  asymptotically, the  effect  of
fluctuations is strongly     enhanced.  The net   result of   spurious
synchronization   is that the     effective  critical  coupling    for
finite-size networks shifts  to lower values   as $N$ decreases. As  a
consequence, the    average Hamming distance   in our  simulations may
vanish even for coupling probabilities  below $q_c$, as illustrated in
Figs. \ref{dt1} and \ref{dt2}.

\begin{figure}
\centering
\resizebox{\columnwidth}{!}{\includegraphics[angle=-90]{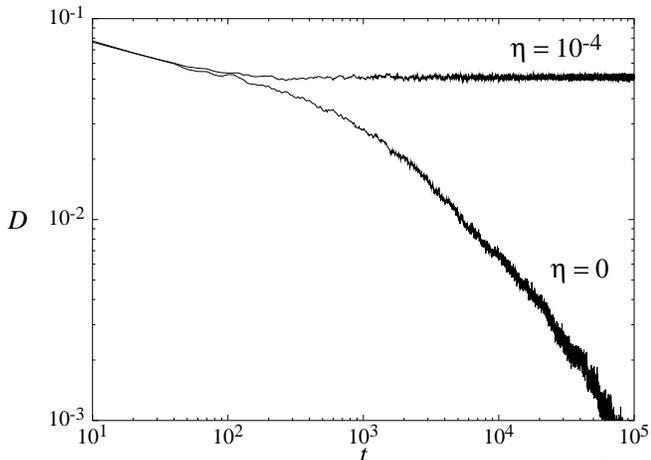}}
\caption{Hamming  distance as   a   function of time  for  $10^3$-site
networks with  $q=0.29$ and two  values of the noise intensity $\eta$,
averaged over   $10^3$     realizations. The  effects   of    spurious
synchronization for $\eta = 0$ are apparent.}
\label{dt3}
\end{figure}

Spurious synchronization can be avoided by adding noise to the system.
Such strategy has already been adopted in this field, specifically, in
the      study      of     globally      coupled     chaotic      maps
\cite{kanekomilnor,manrubia1},  to   prevent synchronization   due  to
round-off errors in computer simulations. We implement the addition of
noise as  a  new  substep in the  dynamics  of  our system. After  the
evolution and coupling substeps, we flip the state of each site in one
of  the networks with  a  small  probability $\eta$. Figure  \ref{dt3}
illustrates  the  effect  of  noise in  the   evolution  of $D(t)$ for
$q=0.29$   and  $N=10^3$.  For  this    coupling intensity, where  the
consequences of spurious synchronization are vast, the behavior of the
Hamming distance with and without noise changes drastically.

Note that noise eliminates spurious  synchronization for $q<q_c$,  but
also  prevents the  KNs to   exactly   synchronize even for   $q>q_c$.
Therefore,    the  critical     behavior   that   characterizes    the
synchronization transition in the absence of noise disappears as noise
is  added,  and is  recovered  only  for  $\eta\to  0$ (but  $\eta\neq
0$). The effect  of noise can be straightforwardly incorporated to the
AM approximation.   

\begin{figure}
\centering
\resizebox{\columnwidth}{!}{\includegraphics[angle=-90]{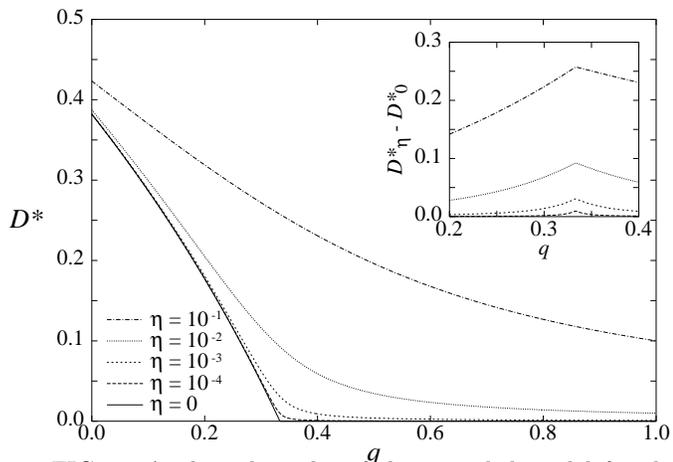}}
\caption{Analytical  results of the annealed  model for the asymptotic
Hamming  distance    $D^*_\eta$,   for different   noise   intensities
$\eta$. The insert shows the deviation $D^*_{\eta} - D^*_{0}$ from the
distance in the absence of noise.}
\label{map2}
\end{figure}

The map that gives the time evolution of the Hamming distance is now
\begin{equation} \label{dnoise}
D(t+1) = (1-\eta) F\left[ D(t) \right] 
+ \eta \left( 1- F\left[ D(t) \right] \right),
\end{equation}
with $F\left[ D(t)  \right]$  given by  Eq. (\ref{ddac}).  As for  the
model without noise, for $K=3$ it is possible to analytically find the
asymptotic distance $D^*_\eta$ predicted by Eq. (\ref{dnoise}).  Figure
\ref{map2} shows $D^*_\eta$ as a  function of the coupling probability
$q$ for  various   noise intensities.  Note the  approximation  to the
critical behavior as  $\eta\to 0$. The insert displays the difference
between $D^*_\eta$ and the asymptotic  distance in the absence of noise
as a function of $q$.

\begin{figure}
\centering
\resizebox{\columnwidth}{!}{\includegraphics[angle=-90]{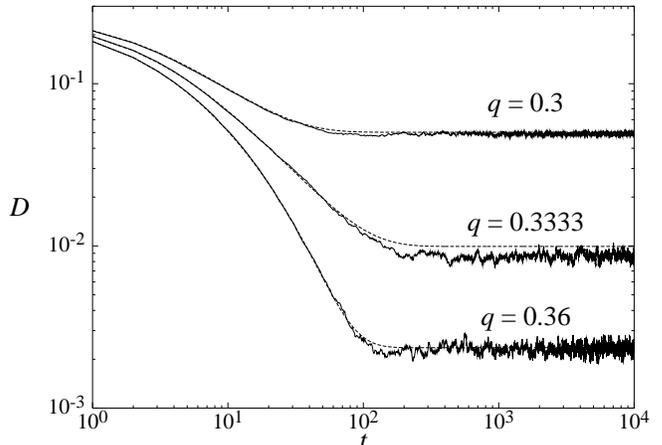}}
\caption{Hamming distance  as a    function of time   for  $10^4$-site
networks and three values  of  the coupling probability $q$,  averaged
over $10^2$ realizations.  The noise intensity is $\eta = 10^{-4}$.}
\label{dt4}
\end{figure}

In   Fig. \ref{dt4}, we compare  the  prediction of Eq. (\ref{dnoise})
with  numerical   results for  the   Hamming  distance of two  coupled
$10^4$-site KNs with noise   intensity  $\eta = 10^{-4}$, for    three
values of  the coupling intensity.  The agreement is  excellent during
the transients,  but   some noticeable  discrepancies  persist in  the
asymptotic value, especially, for $q\approx q_c$.

\begin{figure}
\centering
\resizebox{\columnwidth}{!}{\includegraphics[angle=-90]{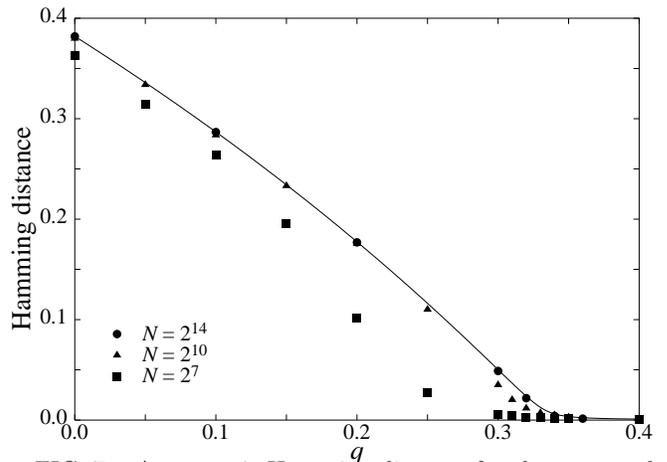}}
\caption{  Asymptotic Hamming distance   for three network  sizes $N$,
averaged over $5 \times 10^3$ time steps and $10^3$ realizations.  The
noise intensity is $\eta =  10^{-4}$. The curve  corresponds to the AM
prediction.}
\label{dq6}
\end{figure}

To study  such discrepancies in detail, and  thus test the AM results,
we compute from our numerical simulations the average asymptotic value
of $D(t)$, defined as
\begin{equation} \label{ave}
\langle D  \rangle =  
{1 \over T} \sum_{t=t_0}^{t_0+T}{D(t)}.
\end{equation}
Averages are performed over  a time span $T$ of  $5 \times 10^3$  time
steps,  when the  asymptotic  regime of  the  coupled system has  been
reached,  i.e. for sufficiently  large  values of $t_0$.  As above, we
choose $\eta=10^{-4}$, and determine  $\langle D \rangle$  for several
values   of  $N$. Results  for  $N=2^7$,  $2^{10}$,  and $2^{14}$  are
presented in Fig.  \ref{dq6}.  The  AM  prediction for  this value  of
$\eta$ is  also   shown, as a curve.    We  see that  the   AM results
systematically overestimate the values of $\langle D\rangle$, and that
the agreement improves for larger values  of $N$. Therefore, even when
noise has been  added to  avoid spurious synchronization,  finite-size
effects persist.

\begin{figure}
\centering
\resizebox{\columnwidth}{!}{\includegraphics[angle=-90]{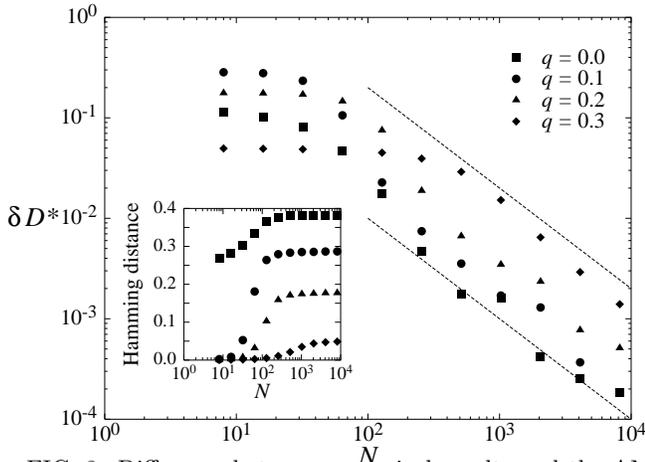}}
\caption{Difference between numerical results   and the AM  prediction
for the asymptotic  Hamming distance as a function  of size,  for four
coupling probabilities $q$. The   dashed lines have  slope  $-1$.  The
insert shows the asymptotic Hamming distance in semilogarithmic scale,
to appreciate the behavior for small values of $N$.}
\label{ndepfig}
\end{figure}

These remanent   finite-size effects  are measured by   the difference
$\delta    D^* = D^{*}_\eta   - \langle   D  \rangle$  between  the AM
prediction  and the numerical average  defined in Eq.  (\ref{ave}). In
Fig.  \ref{ndepfig}  we plot $\delta D^*$  as  a function   of $N$ for
different  coupling intensities.    The insert  shows   $D^*_\eta$ and
$\langle D\rangle$ as  a function of $N$  for the same values  of $q$.
There are two well-differentiated   regimes in the size  dependence of
$\delta D^*$. For small $N$, the deviation between the AM estimate and
numerical results is practically constant. For large values of $N$ the
deviation  decreases,  seemingly  as a  power-law,  $\delta  D^*  \sim
N^{-z}$. Least-square fits for $N>10^2$ yield $z= 1.1 \pm 0.2$ for the 
exponent.

To find an explanation for these finite  size effects, it is necessary
to go beyond the annealed approximation.  In the following, we outline
an approach   to  the calculation  of  the  Hamming distance based  on
statistical averages over states  of the KNs.   The density of a state
$\boldsymbol{\sigma}=(\sigma_1,\dots,\sigma_N)$ of    a single  KN  is
defined as
\begin{equation}
\rho \left( \boldsymbol{\sigma} \right) =
\frac{1}{N} \sum_{i=1}^{N}{\sigma_i}
\end{equation}
[cf. Eq. (\ref{difer})],  whereas the distance between two states
$\boldsymbol{\sigma}$ and $\boldsymbol{\sigma}'$ reads
\begin{equation} \label{D}
D(\boldsymbol{\sigma},\boldsymbol{\sigma}') = 
\frac{1}{N} \sum_{i=1}^{N}{(\sigma_i - \sigma_i')^2}.
\end{equation}

For fixed  $p$, $K$  and  $N$, a  realization $R$   of the network  is
defined by  the connections and the local  rules. We define $\Omega_R$
as the set of all  the states visited by  the KN for this realization,
at asymptotically  large  times and   from  all the possible   initial
conditions.    In other words,  the  set  $\Omega_R$ contains all  the
states that belong to the limit cycles of the dynamics. It is possible
to   introduce   a   probability distribution     ${\cal P}_R   \left(
\boldsymbol{\sigma} \right)$  over $\Omega_R$,  given by the frequency
with  which a  given  state   $\boldsymbol  {\sigma}$ is  visited   at
asymptotically large   times averaged  over  all  initial  conditions.
Averages over $\Omega_R$ will be computed with this distribution.  For
instance, the average of the density $\rho$ is
\begin{equation} \label{dR}
d_R = \langle \rho (\boldsymbol{\sigma}) \rangle _{\Omega_R} =
\sum_{\boldsymbol{\sigma} \in \Omega_R}
{{\cal P}_R (\boldsymbol{\sigma}) \rho (\boldsymbol{\sigma})} .
\end{equation}
The average of the distance between two states, Eq. (\ref{D}), 
can be written as
\begin{equation} \label{Dav}
\langle D(\boldsymbol{\sigma},\boldsymbol{\sigma}') 
\rangle _{\Omega_R}= 2 d_R (1 - d_R) - \frac{2}{N} \sum_{i=1}^{N}
\xi_i^2,
\end{equation}
where $\xi_i  = \langle \sigma_i   \rangle_{\Omega_R}  - d_R=  \langle
\sigma_i' \rangle_{\Omega_R} -  d_R$. Here, we have assumed   that the
occurrence of  states $\boldsymbol{\sigma}$ and $\boldsymbol{\sigma}'$
are probabilistically uncorrelated.

The quantities $\xi_i$ measure, for  each realization of the  network,
the  average deviation  of the state  of   each site from  the average
density  $d_R$  over   the  whole  network.   Let   us introduce   the
distribution $\Gamma_R(\xi )$ as the fraction of sites  in the KN with
deviation $\xi$ for  a specific  realization $R$.  Unfortunately,  the
explicit form of $\Gamma_R(\xi  )$ is not  known.  It is however known
that this  is a   nontrivial  distribution,  in  particular, due    to
existence of the so-called stable core \cite{fly}.  The stable core is
a set of sites  that  have always  the same asymptotic,  fixed states,
irrespectively of  the initial condition.   For  these sites, $\langle
\sigma_i \rangle_{\Omega_R} =0$ or $1$, so that the deviations $\xi_i$
adopt their extremal values, $\xi_i =  -d_R$ or $1-d_R$, respectively.
Using   the  distribution $\Gamma_R   (\xi)$  to  replace  the  sum in
Eq. (\ref{Dav}) by an integral, we have
\begin{equation} \label{q3}
\langle D (\boldsymbol{\sigma},\boldsymbol{\sigma}')
\rangle_{\Omega_R} = 2 d_R (1-d_R) - 
2 \int_{-d_R}^{1-d_R}{\Gamma_R (\xi) \xi^2 d\xi} .
\end{equation}
Now,  analytical results  on  the size of   the stable core \cite{fly}
suggest  that,  as    $N\to  \infty$,  $\Gamma_R(\xi)$  approaches  an
asymptotic profile $\Gamma(\xi)$  which depends on   $p$ and $K$,  but
becomes independent of  the specific realization  of the  network. For
large sizes, we may assume a dependence of the form
\begin{equation} \label{Gamma}
\Gamma_R (\xi) \approx \Gamma (\xi) -N^{-1} \Gamma'_R (\xi),
\end{equation}
where $\Gamma'_R$ is the  first (analytical) correction due  to finite
sizes. Within these assumptions, Eq. (\ref{q3}) takes the form
\begin{equation} \label{D0}
\langle D (\boldsymbol{\sigma},\boldsymbol{\sigma}') 
\rangle_{\Omega_R} = D_0
- \frac{2}{N} \int_{-d_R}^{1-d_R}{\Gamma'_R (\xi) 
\xi^2 d\xi} ,
\end{equation}
where
\begin{equation}
D_0 = 2 d_R (1-d_R) - 2 \int_{-d_R}^{1-d_R}{\Gamma 
(\xi) \xi^2 d\xi} 
\end{equation}
is the asymptotic distance for $N \rightarrow \infty$.

We now  associate $\langle D (\boldsymbol{\sigma}, \boldsymbol{\sigma}
') \rangle_{\Omega_R}$  with the  distance between  the states  of two
coupled KNs at   a given  (long) time.   Indeed,  in  our system  both
networks have the same connections and rules, and correspond therefore
to  the  same  realization $R$   of the   network.    According to the
definition (\ref{difer0})   and (\ref{difer}), after  an average  over
realizations of the network for fixed $p$,  $K$, and $N$ is performed,
the  distance $\langle  D  (\boldsymbol{\sigma}, \boldsymbol{\sigma}')
\rangle_{\Omega_R}$  coincides  with the  Hamming  distance $\langle D
\rangle$, Eq.   (\ref{ave}), considered above.    In Eq.   (\ref{D0}),
thus, $D_0$ should correspond to the Hamming distance predicted by the
AM approximation, valid for $N\to \infty$, and the first correction to
the AM  estimate is given  by the additional  term.  As found from our
simulations, this term  depends on the network  size as $N^{-1}$.  The
ansatz (\ref{Gamma}) is therefore supported by numerical results. Note
that these  conclusions are independent of  the strength  of coupling,
measured by the probability $q$, since the only effect of the coupling
dynamics in connection with the above analysis is to change the set of
asymptotic  states $\Omega_R$  and  the  profile of the   distribution
$\Gamma_R(\xi)$.

\section{Disordered cellular automata and partially ordered networks} 
\label{dca}

According   to the results reported   in  the  previous sections,  the
synchronization transition of  stochastically coupled  KNs on the  one
hand, and of stochastically coupled elementary cellular automata (ECA,
\cite{wolf})  on the other,  are qualitatively different. Namely, they
belong to different universality classes.  Whereas  we have found that
synchronization     in  KNs    appears    through   a    transcritical
bifurcation---with  a critical   exponent   equal to  unity   for  the
asymptotic Hamming  distance---the corresponding transition in ECA has
been shown to exhibit  nontrivial exponents \cite{eca}, which  seem to
suggest     that it belongs  to  the   universality  class of directed
percolation  \cite{grass}. It is  therefore relevant  to study a third
class of systems, intermediate between ECA and generic KNs.

In a generic KN there are two sources of disorder. We have the network
topology,   determined by the  random   choice of connections, and the
local evolution rules,  which are also  chosen at random. On the other
hand, in  cellular automata both the  topology and the dynamical rules
are fully homogeneous. ECA can indeed be interpreted as a very special
subclass of KNs  with $K=3$, where the choice  of  dynamical rules and
connections is  deterministic.   In order to   distinguish between the
effects of disorder  in   the topology  and   in the dynamics  on  the
synchronization transition, we consider now the  subclass of KNs where
the connections are still  chosen at random  but the local rule is the
same for all sites.  We refer to these networks as disordered cellular
automata (DCA).

We focus  here  on the elementary rule   $22$, which exhibits  chaotic
evolution  \cite{wolf}. This rule is  defined  by the Boolean function
$f(\{0,0,1\} )  = f(\{0,1,0\} )=   f(\{1,0,0\} )=1$ and $f=0$ for  the
remaining five possible  neighborhoods (see Table \ref{rule22}). Thus,
the  bias  for rule $22$  is  $p=3/8$.  This DCA, however,  cannot  be
thought of as a KN  with $p=3/8$.  In a generic  KN with this bias, in
fact, the local evolution  could  be substantially different  from the
behavior of  rule $22$. In particular,  the dynamics at some sites may
be governed  by nonchaotic rules,  giving rise to sensible differences
in the global behavior. This  is clearly illustrated, for instance, by
a   measurement of the asymptotic  density  of a  rule-$22$ DCA, which
yields $d \approx 0.423$, instead of $d=p=0.375$.

The formulation of  an annealed  model   of DCA requires  a  different
approach, in order to  account  for the  homogeneity of the  dynamical
rules.  We  define the AM by  reassigning  all the connections at each
time step, but keeping the functions  $f_i=f$ fixed. Suppose that
we have two networks  with overlap $a(t)$.  The probability for a site
to  have  exactly   $K-l$   of   its  reassigned inputs  coming   from
homologous sites in the same state is
\begin{equation}
p_{l}(t)= {K \choose l} a(t)^{K-l}\left[ 1-a(t)\right]^{l} .
\end{equation}
We   introduce  the quantity   $A_{l}$  as  the   probability for  two
homologous sites having all  but  $l$  inputs coming from   homologous
sites in the same state to give the same output.   The overlap at time
$t+1$ is then given by
\begin{equation}
\label{am-dnur}
a(t+1)=\sum\limits_{l=0}^{K}A_{l}p_{l}(t).
\end{equation}
The quantities $A_l$ depend on the evolution  rule, and their value is
fixed.  They can be evaluated  within some approximations, as shown in
the Appendix.  Note that $A_{0}=1$ because, no matter the rule, if the
inputs are all equal the outputs will coincide.  In the annealed model
for KNs  we had $A_{0}=1$ and  $A_{l}=p^{2}+ (1-p)^{2}$ for $l=1,\dots
,K$.  For $K=3$ the map  for the Hamming  distance can be casted  into
the form
\begin{equation} \label{bbb}
D(t+1)  = B_1 D(t) + B_2 D(t)^2 + B_3 D(t)^3,
\end{equation}
with $ B_1= 3 \left( 1-A_1 \right)$, $ B_2= 3  \left( 2 A_1  - A_2 - 1
\right)$, $  B_3= -3A_1 + 3A_2 -A_3  +  1$.  Coupling enters  then the
formulation exactly as in   Eq. (\ref{ddac}), as an  additional factor
$1-q$ in  the evolution of $D(t)$. 

\begin{figure}
\centering
\resizebox{\columnwidth}{!}{\includegraphics[angle=-90]{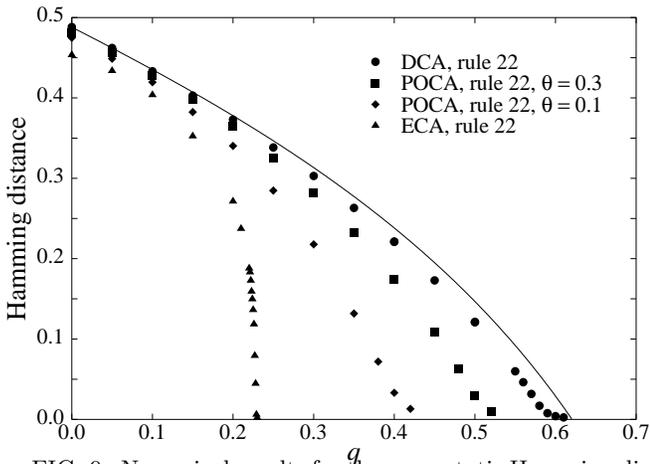}}
\caption{Numerical  results  for the   asymptotic Hamming distance  in
coupled $2^{11}$-site  disordered cellular automata with the evolution
rule $22$, for different amounts of disorder. The curve stands for the
annealed model prediction for DCA. }
\label{figdca}
\end{figure}

In Fig. \ref{figdca}, the curve stands for  the asymptotic value $D^*$
as   a function   of  the  coupling  probability   predicted from  Eq.
(\ref{bbb}) for rule $22$.  The prediction is qualitatively similar to
that  for  KNs,  in   particular,   in the    region   close to    the
synchronization  transition. Numerical results on  DCA  with rule $22$
for $N=2^{11}$  are  also  shown  in  Fig.    \ref{figdca}.   To avoid
spurious synchronization, a small amount of noise, $\eta=10^{-4}$, has
been added. The agreement with the  AM is reasonably good, though some
systematic deviations    are  clearly  visible  in    the zone of  the
transition.  As   before, these  deviations    may be   attributed  to
finite-size effects.

Finally, as an interpolation between DCA and ECA we consider partially
ordered topologies (POCA), constructed in the following way.  We start
with an $N$-site ECA, which  consists of a one-dimensional array where
each site  is connected to  itself  and to its  two nearest neighbors.
Then, for each site, each neighbor is redrawn at random from the whole
network   with    probability  $\theta$.   With     the  complementary
probability, $1-\theta$, the    neighbor is preserved.  The  resulting
topological  structure  of   connections   is analogous  to    that of
small-world networks \cite{swm}.  For $\theta = 1$, we recover the DCA
networks discussed  above.  Results of  numerical simulations  on POCA
with rule $22$  are shown in  Fig.  \ref{figdca} for  $\theta=0.1$ and
$\theta=0.3$. Data for ECA with rule $22$ ($\theta=0$) are also shown.
We see that,  in spite of the relatively  low values of  $\theta$, the
asymptotic    Hamming   distance of  POCA   depends   on  the coupling
probability  $q$ in a way qualitatively  similar to  that  of KNs.  In
particular, it does not exhibit the abrupt  dependence on $q$ observed
in  ECA near the critical  point. This would  be in agreement with the
crossover scenario  in small-world networks \cite{crossover}, where it
is  known that even small  amounts  of disorder induce behaviors which
already resemble that of fully random structures.

\section{Summary and discussion} \label{summary}

We have  studied the  behavior  of  two Kauffman networks  interacting
through a form of symmetric  stochastic  coupling.  As for many  other
localized or    extended, random or  deterministic,  coupled dynamical
systems
\cite{logmaps,rossler,hamilton,neural,amengual,eca,bagnoli,grass},  we
have  found   that coupled Kauffman   networks   can synchronize their
evolution if coupling  is strong  enough.  In our   case,  there is  a
critical  value of the coupling probability  $q$  beyond which the two
networks converge to the same trajectory as time elapses.

In contrast with the situation encountered  for other extended systems
\cite{neural,amengual,eca,bagnoli,grass,batistutacitadinoonorario},
however, for   Kauffman  networks it  has  been  possible  to give  an
analytical description of the synchronization transition, in excellent
agreement  with  numerical  results for   large-size  systems.    This
formulation   is provided by an  extension  of  the so-called annealed
model  \cite{d+p,d+w} to the  system   of coupled networks.  The model
gives  the evolution of the overlap  between the two networks---or, in
other words,  of  their Hamming  distance---and  makes it possible  to
evaluate its asymptotic value.  The asymptotic Hamming  distance $D^*$
is used as an order parameter  for the synchronization transition. For
coupled Kauffman networks  in  the chaotic  phase, the annealed  model
predicts the existence of a critical  coupling probability $q_c$, such
that $D^*$  is  finite for $q<q_c$ and  vanishes  for  $q>q_c$. At the
critical   point,  $D^* \sim   |q-q_c|$,  and  the transition  has the
character of a transcritical bifurcation.

The transition  predicted    by the annealed model   is  qualitatively
different from that observed in the deterministic version of this kind
of  systems, more specifically, in  cellular  automata. For elementary
cellular   automata,  in fact,  the  critical  behavior of the Hamming
distance  exhibits   nontrivial exponents   \cite{eca}.   It has  been
suggested that, at least for some evolution rules, the synchronization
transition in cellular  automata  belongs  to the class   of  directed
percolation  \cite{grass}. On the  other  hand,  by analogy with   the
problem of  damage  spreading, synchronization   in Kauffman  networks
could  belong to the    class of directed  percolation in   disordered
systems \cite{grass2}.

Results  from extensive    numerical simulations  of  relatively large
networks ($10^4$ sites),   performed  for  Kauffman networks   in  the
chaotic  regime ($p=1/2$,   $K=3$),  are in  good  agreement with  the
predictions of the annealed model  both in the temporal evolution  and
in the asymptotic behavior of  the system. However, for small networks
some systematic departures from  the analytical results  are apparent.
These deviations can  be partially explained  taking into account  the
occasional events of spurious synchronization for  $q<q_c$, due to the
effect of suitably  large   fluctuations on our discrete   finite-size
system. The effect of these fluctuations becomes more important as the
coupling probability approaches  its   critical value.  In  any  case,
spurious  synchronization can be  successfully  eliminated by adding a
small  amount of noise  to the dynamics.   Moreover, noise can also be
encompassed into the  annealed model. We  have introduced noise in the
simulations     and observed    that   remanent   finite-size  effects
persist.  These  must now be ascribed  to  the failure of the annealed
model in  describing finite  networks.  By  means of  a  more detailed
statistical description of the  overlaps between networks, in fact, we
have been able to account  for such remanent effects, also  explaining
the dependence of the deviation from the annealed model on the network
size.

Finally, we have presented preliminary numerical results on disordered
and partially  ordered stochastically coupled cellular automata. These
systems   can be  seen as providing   a kind  of interpolation between
Kauffman networks, with    their completely   random  connections  and
dynamical rules,  and cellular automata,  which are fully ordered.  In
the case of  disordered networks with the same  evolution rule on  all
the sites, it is possible to extend the annealed model, which predicts
the same class of synchronization transition as for Kauffman networks.
Numerical simulations  agree with  these predictions.  Increasing  the
order in  the  connections by  means of  a  scheme  analogous to   the
construction  of   a small-world  network  \cite{swm},  we   have also
considered partially disordered structures.  Even for small amounts of
structural disorder, the  behavior associated with the synchronization
transition seems to resemble that of  Kauffman networks more than that
of cellular     automata.  This leads   us   to  conjecture   that the
synchronization transition of partially disordered  automata is in the
same universality    class  as    for   coupled  Kauffman    networks.
Nevertheless,   further   extensive   simulations    and   a   careful
determination  of the critical  exponents  are  necessary to draw  any
conclusions on this point.

\section*{Acknowledgment}

This work has been partially carried out at the Abdus Salam Centre for
Theoretical Physics (Trieste, Italy). The authors thank the Centre for
hospitality.

\appendix

\section*{Coefficients for the annealed model}

In this  Appendix we  illustrate  the computation of the  coefficients
$A_l$ in Eq. (\ref{am-dnur}) with an explicit  example. We recall that
$A_l$ is defined  as the probability  for two homologous sites  having
all but $l$  inputs coming from homologous  sites in the same state to
give the same output.

Let   us assume that the   frequency with which   a given neighborhood
appears in  a state of  the whole network  is fully  determined by the
density $d$ of the state.  Namely, we neglect the correlations between
neighborhood   frequencies,  associated  with   the  spatial  patterns
generated   by  the dynamics.  For   $K=3$,   there are eight possible
neighborhoods,  which  we  label from  $0$ to  $7$  as shown  in Table
\ref{rule22}. Within the above assumption, the frequency $p_i$ of each
neighborhood $i$ can be estimated in terms of the density $d$ as
\begin{equation} \label{a1}
\begin{array}{l}
p_0= (1-d)^3,  \\ 
p_1= p_2=p_4= d (1-d)^2 , \\ 
p_3= p_5=p_6= d^2 (1-d) , \\ 
p_7= d^3. 
\end{array}
\end{equation}
Moreover, we note that the density $d$ can in turn be written in terms
of the frequencies  $p_i$  of neighborhoods with nonzero  output.  For
rule $22$  these    neighborhoods are  $\{1,0,0\}$,  $\{0,1,0\}$   and
$\{0,0,1\}$ (see Table \ref{rule22}), so that we have
\begin{equation} \label{a2}
d = p_1 + p_2 + p_4.
\end{equation}
Combining Eqs. (\ref{a1}) and (\ref{a2}) yields $d=1-\sqrt{3}/3\approx 
0.423$, which  agrees with the numerical measurement reported in Sect. 
\ref{dca}.  The corresponding  values of the   frequencies  $p_i$  are
shown in Table \ref{rule22}. They are  in very good agreement with the
values  obtained   from  numerical simulations,    also  shown in  the
table. This supports our above assumption of uncorrelated neighborhood
frequencies.

Once the frequencies  $p_i$ are known,  we are  able to calculate  the
coefficients $A_l$.   As  a specific example,  we   discuss $A_3$.  By
definition, this is the  probability  for two homologous sites  having
all but $3$ inputs coming from  homologous sites in  the same state to
give  the  same  output. Thus,  we  are  in  the case  where   the two
neighborhoods have all the homologous sites in  different states.  The
pairs of neighborhoods    that satisfy this  condition  are $\{0,7\}$,
$\{1,6\}$, $\{2,5\}$,  and $\{3,4\}$.  Among  them,  however, only the
pair $\{0,7\}$ has the  same output for  both neighborhoods (see Table
\ref{rule22}). The coefficient $A_3$ is therefore given by
\begin{equation}
A_3 = \frac{p_0 p_7}{p_0 p_7 + p_1 p_6 + p_2 p_5 + p_3 p_4} 
=\frac{1}{4}.
\end{equation}
The computation of the other coefficients is accomplished in a similar 
way, yielding $A_1=4(19-8\sqrt{3})/169$ and $A_2=(9+\sqrt{3})/13$. 
Moreover, $A_0=1$.

\begin{table}
\begin{tabular}{ccccc}
label & neighborhood &  output & 
$p_{\rm numerical}$ & $p_{\rm analytical}$ \\
\tableline
0 & $\{0,0,0\}$ & 0 & 0.19350 & 0.19294 \\
1 & $\{0,0,1\}$ & 1 & 0.14075 & 0.14096 \\
2 & $\{0,1,0\}$ & 1 & 0.14071 & 0.14096 \\
3 & $\{0,1,1\}$ & 0 & 0.10287 & 0.10298 \\
4 & $\{1,0,0\}$ & 1 & 0.14079 & 0.14096 \\
5 & $\{1,0,1\}$ & 0 & 0.10290 & 0.10298 \\
6 & $\{1,1,0\}$ & 0 & 0.10289 & 0.10298 \\
7 & $\{1,1,1\}$ & 0 & 0.07559 & 0.07524 \\
\end{tabular}
\caption{  Cellular automata neighborhoods  and  their outputs for the
evolution  rule  $22$.  The frequencies   $p$   for each  neighborhood
obtained from numerical results and  from the analytical approximation
used in the Appendix are also quoted.}
\label{rule22}
\end{table}

\end{document}